%% file: main.tex
\pdfoutput=1

\documentclass[12pt,a4paper]{article}

\usepackage{ifthen} 
\newboolean{pdflatex}
\setboolean{pdflatex}{true} 

\newboolean{articletitles}
\setboolean{articletitles}{true} 

\newboolean{uprightparticles}
\setboolean{uprightparticles}{false} 

\def\paperauthors{LHCb collaboration} 
\def\paperasciititle{Amplitude analysis of B KKpi decays} 
\def\papertitle{Amplitude analysis of \kkpi decays} 
\def\paperkeywords{{High Energy Physics}, {LHCb}} 
\def\papercopyright{\the\year\ CERN for the benefit of the LHCb collaboration} 
\def\paperlicence{CC-BY-4.0 licence}
\def\paperlicenceurl{https://creativecommons.org/licenses/by/4.0/}

\input{preamble}
\usepackage{longtable} 
\begin{document}

\renewcommand{\thefootnote}{\fnsymbol{footnote}}
\setcounter{footnote}{1}

\input{title-LHCb-PAPER}


\renewcommand{\thefootnote}{\arabic{footnote}}
\setcounter{footnote}{0}



\pagestyle{plain} 
\setcounter{page}{1}
\pagenumbering{arabic}


%

\input{introduction}

\input{detector3}

\input{selection}

\input{models}

\input{results}

\input{conclusions}


\input{acknowledgements}

\addcontentsline{toc}{section}{References}
\bibliographystyle{LHCb}
 \bibliography{main.bbl} 
 
\newpage
\input{LHCb_Authorship_13-Dec-2018.tex}

\end{document}

%% file: preamble.tex
\usepackage[top=1in, bottom=1.25in, left=1in, right=1in]{geometry}

\usepackage{microtype}
\usepackage{lineno}  
\usepackage{xspace} 
\usepackage{caption} 

\usepackage{graphicx}  
\usepackage{color}
\usepackage{colortbl}
\graphicspath{{./figs/}} 
\DeclareGraphicsExtensions{.pdf,.PDF,png,.PNG}

\usepackage{amsmath} 
\usepackage{amssymb}
\usepackage{amsfonts}
\usepackage{upgreek} 

\newcommand*\patchAmsMathEnvironmentForLineno[1]{%
\expandafter\let\csname old#1\expandafter\endcsname\csname #1\endcsname
\expandafter\let\csname oldend#1\expandafter\endcsname\csname
end#1\endcsname
 \renewenvironment{#1}%
   {\linenomath\csname old#1\endcsname}%
   {\csname oldend#1\endcsname\endlinenomath}%
}
\newcommand*\patchBothAmsMathEnvironmentsForLineno[1]{%
  \patchAmsMathEnvironmentForLineno{#1}%
  \patchAmsMathEnvironmentForLineno{#1*}%
}
\AtBeginDocument{%
\patchBothAmsMathEnvironmentsForLineno{equation}%
\patchBothAmsMathEnvironmentsForLineno{align}%
\patchBothAmsMathEnvironmentsForLineno{flalign}%
\patchBothAmsMathEnvironmentsForLineno{alignat}%
\patchBothAmsMathEnvironmentsForLineno{gather}%
\patchBothAmsMathEnvironmentsForLineno{multline}%
\patchBothAmsMathEnvironmentsForLineno{eqnarray}%
}

\usepackage{hyperxmp}

\usepackage[pdftex, pdfauthor={\paperauthors}, pdftitle={\paperasciititle}, pdfkeywords={\paperkeywords}, pdfcopyright={Copyright (C) \papercopyright}, pdflicenseurl={\paperlicenceurl}]{hyperref}

\usepackage[colorinlistoftodos,textsize=scriptsize]{todonotes}

\usepackage[all]{hypcap} 

\input{lhcb-symbols-def} 

\usepackage{overpic}
\usepackage{float}
\usepackage{adjustbox}
\usepackage{graphicx}
\usepackage{siunitx}
\usepackage{multirow}

\usepackage{cite} 
\usepackage{mciteplus}

%% file: lhcb-symbols-def.tex
\usepackage{xspace} 
\usepackage{upgreek}

\def\hhh {\ensuremath{{\Bpm \to h^{\pm} h^{ +} h^{-}}}\xspace}

\def\pipipi {\ensuremath{{\Bpm \to \pipm \pip \pim}}\xspace}
\def\kpipi {\ensuremath{{\Bpm \to \Kpm \pip \pim}}\xspace}

\def\kkpi {\ensuremath{{\Bpm \to \pipm \Kp \Km }}\xspace}

\def\kkpip {\ensuremath{{\Bp \to \pip \Km \Kp }}\xspace}
\def\kkpim {\ensuremath{{\Bm \to \pim \Kp \Km}}\xspace}

\def\mmpik {\ensuremath{m^2_{\pipm\!\Kmp }}\xspace}                    

\def\mmkk {\ensuremath{m^2_{\Kp \!\Km}}\xspace}                            
\def\mkk {\ensuremath{m({\Kp \Km})}\xspace}                            

\def\lhcb   {\mbox{LHCb}\xspace}

\def\babar  {\mbox{BaBar}\xspace}

\def\MagUp {\mbox{\em Mag\kern -0.05em Up}\xspace}

\ifthenelse{\boolean{uprightparticles}}%
{

 \def\Ppi         {\ensuremath{\uppi}\xspace}

 \def\PDelta      {\ensuremath{\Delta}\xspace}                 
 \def\PXi         {\ensuremath{\Xi}\xspace}                 
 \def\PLambda     {\ensuremath{\Lambda}\xspace}                 
 \def\PSigma      {\ensuremath{\Sigma}\xspace}                 
 \def\POmega      {\ensuremath{\Omega}\xspace}                 
 \def\PUpsilon    {\ensuremath{\Upsilon}\xspace}

 \def\PB      {\ensuremath{\mathrm{B}}\xspace}                 
                  
 \def\PD      {\ensuremath{\mathrm{D}}\xspace}

 \def\PK      {\ensuremath{\mathrm{K}}\xspace}

 \def\Pb      {\ensuremath{\mathrm{b}}\xspace}                 
 \def\Pc      {\ensuremath{\mathrm{c}}\xspace}

 \def\Pi      {\ensuremath{\mathrm{i}}\xspace}

 \def\thebaroffset{0.0em}
}
{

 \def\Ppi         {\ensuremath{\pi}\xspace}

 \mathchardef\PDelta="7101
 \mathchardef\PXi="7104
 \mathchardef\PLambda="7103
 \mathchardef\PSigma="7106
 \mathchardef\POmega="710A
 \mathchardef\PUpsilon="7107
                  
 \def\PB      {\ensuremath{B}\xspace}                 
                  
 \def\PD      {\ensuremath{D}\xspace}

 \def\PK      {\ensuremath{K}\xspace}

 \def\Pb      {\ensuremath{b}\xspace}                 
 \def\Pc      {\ensuremath{c}\xspace}

 \def\Pi      {\ensuremath{i}\xspace}

 \def\thebaroffset{0.18em}
}
\newcommand{\offsetoverline}[2][\thebaroffset]{\kern #1\overline{\kern -#1 #2}}%

\makeatletter
\ifcase \@ptsize \relax
  \newcommand{\miniscule}{\@setfontsize\miniscule{4}{5}}
\or
  \newcommand{\miniscule}{\@setfontsize\miniscule{5}{6}}
\or
  \newcommand{\miniscule}{\@setfontsize\miniscule{5}{6}}
\fi
\makeatother

\DeclareRobustCommand{\optbar}[1]{\shortstack{{\miniscule (\rule[.5ex]{1.25em}{.18mm})}
  \\ [-.7ex] $#1$}}

\def\cquark    {{\ensuremath{\Pc}}\xspace}

\def\bquark    {{\ensuremath{\Pb}}\xspace}

\def\pion   {{\ensuremath{\Ppi}}\xspace}

\def\pip    {{\ensuremath{\pion^+}}\xspace}
\def\pim    {{\ensuremath{\pion^-}}\xspace}
\def\pipm   {{\ensuremath{\pion^\pm}}\xspace}

\def\kaon    {{\ensuremath{\PK}}\xspace}

\def\KorKbar {\kern \thebaroffset\optbar{\kern -\thebaroffset \PK}{}\xspace}

\def\Kp      {{\ensuremath{\kaon^+}}\xspace}
\def\Km      {{\ensuremath{\kaon^-}}\xspace}
\def\Kpm     {{\ensuremath{\kaon^\pm}}\xspace}
\def\Kmp     {{\ensuremath{\kaon^\mp}}\xspace}

\def\DorDbar {\kern \thebaroffset\optbar{\kern -\thebaroffset \PD}\xspace}

\def\B       {{\ensuremath{\PB}}\xspace}

\def\BorBbar {\kern \thebaroffset\optbar{\kern -\thebaroffset \PB}\xspace}

\def\Bu      {{\ensuremath{\B^+}}\xspace}
\def\Bub     {{\ensuremath{\B^-}}\xspace}
\def\Bp      {{\ensuremath{\Bu}}\xspace}
\def\Bm      {{\ensuremath{\Bub}}\xspace}
\def\Bpm     {{\ensuremath{\B^\pm}}\xspace}

\def\Y#1S{\ensuremath{\PUpsilon{(#1S)}}\xspace}

\def\LorLbar     {\kern \thebaroffset\optbar{\kern -\thebaroffset \PLambda}\xspace}

\def\to                 {\ensuremath{\rightarrow}\xspace}

\def\CP                {{\ensuremath{C\!P}}\xspace}

\def\CPV
{{\ensuremath{C\!PV}}\xspace}

\def\AT#1     {\ensuremath{A_{\mathrm{T}}^{#1}}\xspace}           

\def\C#1      {\ensuremath{\mathcal{C}_{#1}}\xspace}                       
\def\Cp#1     {\ensuremath{\mathcal{C}_{#1}^{'}}\xspace}                    
\def\Ceff#1   {\ensuremath{\mathcal{C}_{#1}^{\mathrm{(eff)}}}\xspace}        
\def\Cpeff#1  {\ensuremath{\mathcal{C}_{#1}^{'\mathrm{(eff)}}}\xspace}       
\def\Ope#1    {\ensuremath{\mathcal{O}_{#1}}\xspace}                       
\def\Opep#1   {\ensuremath{\mathcal{O}_{#1}^{'}}\xspace}                    

\newcommand{\aunit}[1]{\ensuremath{\text{\,#1}}}       

\newcommand{\tev}{\aunit{Te\kern -0.1em V}\xspace}
\newcommand{\gev}{\aunit{Ge\kern -0.1em V}\xspace}
\newcommand{\mev}{\aunit{Me\kern -0.1em V}\xspace}
\newcommand{\kev}{\aunit{ke\kern -0.1em V}\xspace}
\newcommand{\ev}{\aunit{e\kern -0.1em V}\xspace}
\newcommand{\mevc}{\ensuremath{\aunit{Me\kern -0.1em V\!/}c}\xspace}
\newcommand{\gevc}{\ensuremath{\aunit{Ge\kern -0.1em V\!/}c}\xspace}
\newcommand{\mevcc}{\ensuremath{\aunit{Me\kern -0.1em V\!/}c^2}\xspace}
\newcommand{\gevcc}{\ensuremath{\aunit{Ge\kern -0.1em V\!/}c^2}\xspace}
\newcommand{\gevgevcccc}{\ensuremath{\gev^2/c^4}\xspace} 

\def\fm   {\aunit{fm}\xspace}

\def\fb   {\ensuremath{\aunit{fb}}\xspace}
\def\invfb   {\ensuremath{\fb^{-1}}\xspace}

\def\gsim{{~\raise.15em\hbox{$>$}\kern-.85em
          \lower.35em\hbox{$\sim$}~}\xspace}
\def\lsim{{~\raise.15em\hbox{$<$}\kern-.85em
          \lower.35em\hbox{$\sim$}~}\xspace}

\def\degrees{\ensuremath{^{\circ}}\xspace}

\def\evtgen     {\mbox{\textsc{EvtGen}}\xspace}

\def\geant      {\mbox{\textsc{Geant4}}\xspace}

\def\photos     {\mbox{\textsc{Photos}}\xspace}

\def\pythia     {\mbox{\textsc{Pythia}}\xspace}

\xspace

\def\tell1  {TELL1\xspace}
\def\ukl1   {UKL1\xspace}



%% file: title-LHCb-PAPER.tex

\begin{titlepage}
\pagenumbering{roman}

\vspace*{-1.5cm}
\centerline{\large EUROPEAN ORGANIZATION FOR NUCLEAR RESEARCH (CERN)}
\vspace*{1.5cm}
\noindent
\begin{tabular*}{\linewidth}{lc@{\extracolsep{\fill}}r@{\extracolsep{0pt}}}
\ifthenelse{\boolean{pdflatex}}
{\vspace*{-1.5cm}\mbox{\!\!\!\includegraphics[width=.14\textwidth]{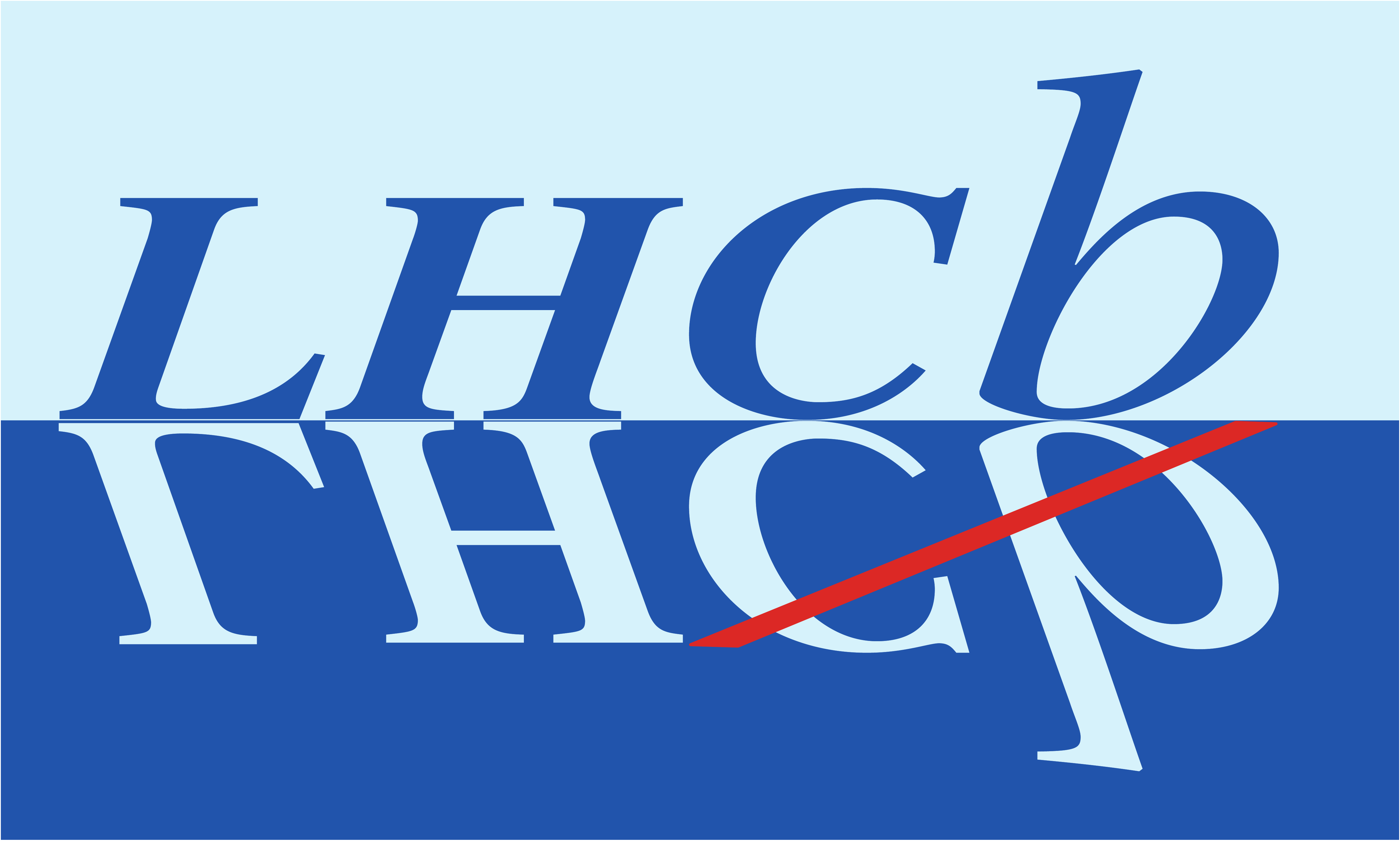}} & &}%
{\vspace*{-1.2cm}\mbox{\!\!\!\includegraphics[width=.12\textwidth]{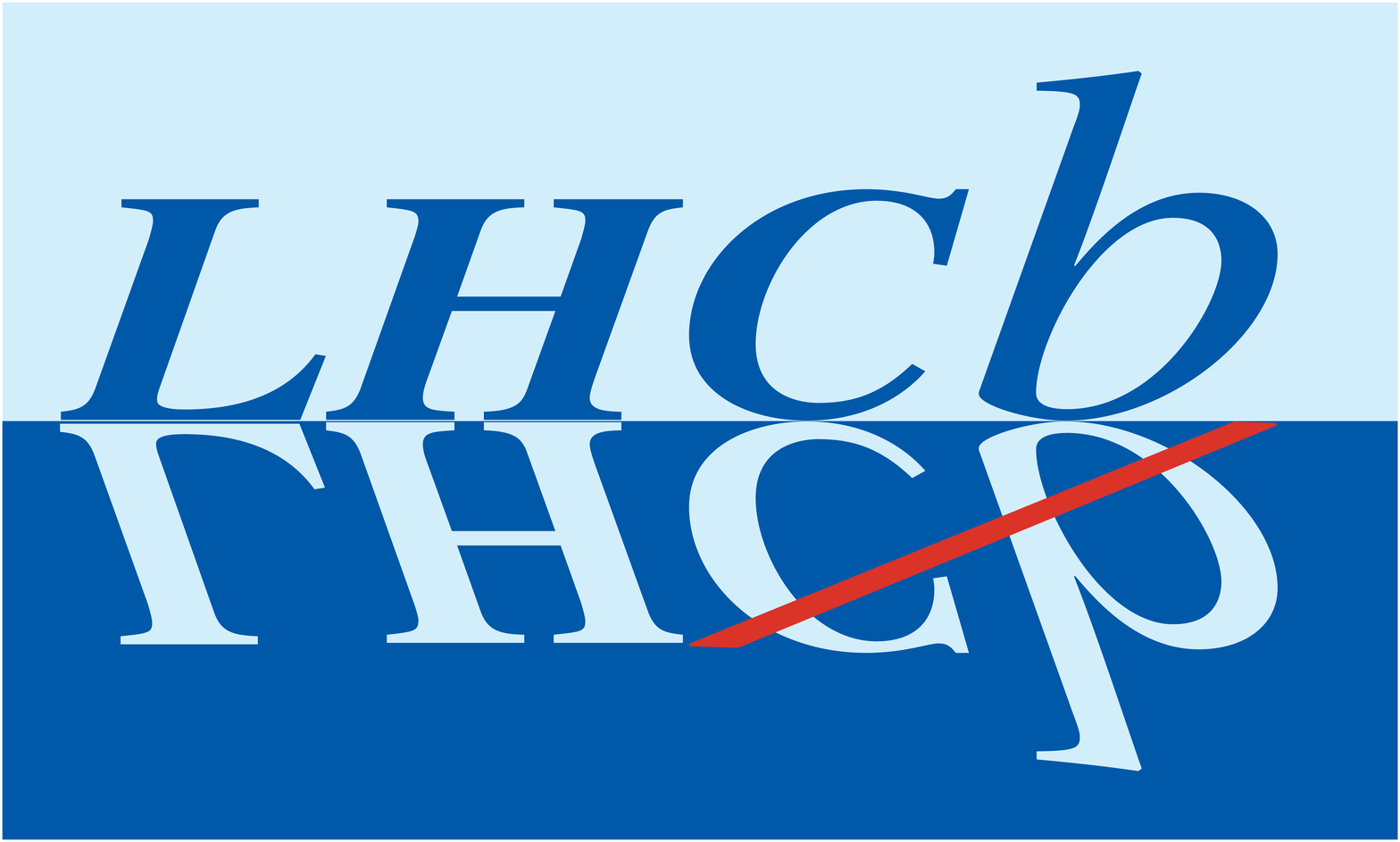}} & &}%
\\
 & & CERN-EP-2019-062 \\  
 & & LHCb-PAPER-2018-051 \\  
 & & December 6, 2019 \\ 
 & & \\
\end{tabular*}

\vspace*{4.0cm}

{\normalfont\bfseries\boldmath\huge
\begin{center}
  \papertitle 
\end{center}
}

\vspace*{2.0cm}

\begin{center}
\paperauthors\footnote{Authors are listed at the end of this paper.}
\end{center}

\vspace{\fill}

\begin{abstract}

%
%
  \noindent The first amplitude analysis of the \kkpi decay is reported based on a data sample corresponding to an integrated luminosity of 3.0\invfb of $pp$ collisions recorded in 2011 and 2012  with the \lhcb detector. The data are found to be best described by a coherent sum of five resonant structures plus a nonresonant component and a contribution from $\pi \pi \leftrightarrow  KK$ $S$-wave rescattering. The dominant contributions in the \pipm\Kmp and \Kp\Km systems are the nonresonant  and  the $B^\pm \to \rho(1450)^{0}\pi^\pm$ amplitudes, respectively,  with fit fractions around 30\%. For the rescattering contribution, a sizeable fit fraction is observed. This component has the largest \CP asymmetry reported to date for a single amplitude of $(-66 \pm 4 \pm 2)\%$, where the first uncertainty is statistical and the second systematic. No significant \CP violation is observed in the other contributions.  
   
\end{abstract}

\vspace*{2.0cm}

\begin{center}
  Published in Phys. Rev. Lett. 123 (2019) 231802  
\end{center}

\vspace{\fill}

{\footnotesize 
\centerline{\copyright~\papercopyright. \href{\paperlicenceurl}{\paperlicence}.}}
\vspace*{2mm}

\end{titlepage}


\newpage
\setcounter{page}{2}
\mbox{~}
%
%
%
%

\cleardoublepage

%% file: introduction.tex

Charge-parity (\CP) symmetry is known to be broken in
weak interactions. In two-body charged \B-meson decays, the only \CP-violating observable is the difference of the partial decay widths for particle and anti-particle over their sum. For three- and multi-body processes, the decay dynamics is very rich, thanks to possible interfering intermediate resonant and nonresonant amplitudes, and therefore \CP violation (\CPV) can be manifested as charge asymmetries that may vary and even change sign throughout the different regions of the observed phase space.

Several experiments have reported sizeable localised \CP asymmetries in the phase space of charmless three-body \Bpm decays~\cite{LHCb-PAPER-2014-044, Hsu:2017kir, Aubert:2007xb, Aubert:2009av, Lees:2012kxa, Aubert:2008bj, Garmash:2005rv}. The \pipipi and \kkpi decays, having the same flavour quantum numbers, are coupled by final-state strong interactions, in particular through the rescattering process $\pip\pim \leftrightarrow \Kp\Km$. The \pipipi decay, with three times  larger branching fraction, may proceed through resonances from the $b \to u$ ($\bar b \to \bar u)$ tree transitions as well as from $b \to d$ ($\bar b \to \bar d)$ loop-induced penguin processes. On the other hand, the production of resonances in the \kkpi decay is limited: $\pi^\pm K^\mp$  resonances can only be obtained from penguin transitions; $K^+K^-$ resonances can come from tree-level transitions, but with the $s\bar{s}$ contribution highly suppressed by the OZI rule~\cite{Okubo:1963fa, Zweig:352337, Iizuka:1966fk}. In the \kkpi decay, no significant $\phi(1020)\to\Kp\Km$ contribution has been seen~\cite{LHCb-PAPER-2013-048}. However, a concentration of events is observed just above the $\phi(1020)$ region in the \Kp\Km invariant-mass spectrum. This corresponds to the region where the $S$-wave  $\pip\pim \leftrightarrow \Kp\Km$ rescattering effect is seen, as shown by elastic scattering experiments~\cite{Estabrooks:1973zd,Cohen:1980cq}. Intriguingly, in this same region, large \CP asymmetries have been observed~\cite{LHCb-PAPER-2014-044,Nogueira:2015tsa}. As proposed in Refs.~\cite{Wolfenstein:1990ks,Bigi:2000yz}, this could be a manifestation of \CPV arising from amplitudes with different rescattering strong phases as well as different weak phases. 
 
 A better understanding of the \CPV mechanisms occurring in three-body hadronic $B$ decays can be achieved through full amplitude analyses. In this Letter, the first amplitude analysis of the decay \kkpi is performed based on a data sample corresponding to an integrated luminosity of 3.0\invfb collected in 2011 and 2012.  The isobar model formalism~\cite{Fleming:1964zz,Herndon:1973yn}, which assumes that the total decay amplitude is a coherent sum of  intermediate two-body states, is applied. A rescattering amplitude is also included. The magnitudes and phases of the coupling to intermediate states are determined independently for $\Bp\to \pip\Km\Kp$ and $\Bm\to\pim\Kp\Km$ decays, allowing for \CP violation.

%% file: detector3.tex
The \lhcb detector is a single-arm forward spectrometer covering the \mbox{pseudorapidity} range $2<\eta<5$ equipped with charged-hadron identification detectors, calorimeters, and muon detectors; and it is designed for the study of particles containing \bquark or \cquark quarks~\cite{LHCb-DP-2008-001,LHCb-DP-2014-002}.

Simulated samples, needed to determine the signal efficiency as well as for background studies, are generated using \pythia~\cite{Sjostrand:2007gs} with a specific \lhcb configuration~\cite{LHCb-PROC-2010-056}. Decays of hadronic particles
are described by \evtgen~\cite{Lange:2001uf}, in which final-state
radiation is generated using \photos~\cite{Golonka:2005pn}. The
interaction of the generated particles with the detector and its
response are implemented using the \geant toolkit~\cite{Allison:2006ve, *Agostinelli:2002hh} as described in
Ref.~\cite{LHCb-PROC-2011-006}.

%% file: selection.tex

 In a preselection stage, \Bpm candidates are reconstructed by requiring three charged tracks forming a good-quality secondary vertex, with loose requirements imposed on their momentum, transverse momentum and impact parameter with respect to any primary vertex. The momentum vector of the $B$ candidate should point back to a primary vertex, from which the \Bpm vertex has to be significantly separated. To remove contributions from charm decays, candidates for which the two-body invariant masses  $m(K^{\pm} \pi^{\mp})$ and $m(K^+ K^-)$ are within $30\mevcc$ of the known value of the $D^0$ mass~\cite{PDG2018} are excluded. 

A multivariate selection based on a boosted decision tree (BDT) algorithm~\cite{Breiman, Roe:2004na} is applied to reduce the combinatorial background (random combination of tracks). The BDT is described in Ref.~\cite{LHCb-PAPER-2014-044};  it is trained using a combination of \hhh samples of simulated events (where $h$ can be either a pion or a kaon) as signal, and data in the  high-mass region $5.40 < m(\pi^{\pm} \pi^+ \pi^-) < 5.58\gevcc$ of a \pipipi sample as background. The \pipipi sample is used as a proxy for the combinatorial background because, among the various \hhh channels, it is the only one whose high mass region is populated just by combinatorial background. The selection requirement on the BDT response is chosen to maximize the ratio $N_S/\sqrt{N_S + N_B}$, where $N_S$ and $N_B$ represent the expected number of signal and background candidates in data, respectively, within an invariant mass window of approximately $40\mevcc$ around the \Bpm mass in the data~\cite{LHCb-PAPER-2014-044}.

Particle identification criteria are used to reduce the crossfeed from other $b$-hadron decays, in particular {\bf $K \leftrightarrow \pi$} misidentification. Muons are rejected by a veto applied to each track~\cite{LHCb-DP-2013-001}. Events with more than one candidate are discarded. 

An unbinned extended maximum-likelihood fit is applied simultaneously to the \pip\Km\Kp and \pim\Kp\Km mass spectra in order to obtain the total signal yields and the raw asymmetry, defined as the difference of \Bm and \Bp signal yields  divided by their sum. Three types of background sources are identified: the residual combinatorial background,  partially reconstructed decays (mostly from four-body decays) and crossfeed from other $B$-meson decays. The parametrisation of crossfeed and partially reconstructed backgrounds is performed using simulated samples that satisfy the same selection criteria as the data. 
 From the result of the fit, yields
for signal and background sources are obtained \cite{LHCb-PAPER-2014-044}. 

Candidates within the mass region $5.266 < m(\pi^{\pm} K^+ K^-) < 5.300$\,\gevcc, referred to as the signal region, are used for the amplitude analysis. This region contains 2052 $\pm$ 102 (1566 $\pm$ 84) of \Bp(\Bm) signal candidates. The relative contribution from the combinatorial background is 23\%, with a charge asymmetry compatible with zero within one standard deviation. The main crossfeed contamination comes from \kpipi decays which contribute in 2.7\% with a charge asymmetry of 2.5\%~\cite{LHCb-PAPER-2014-044}. Another 0.6\% comes from $\phi(1020)$ mesons randomly associated with a pion, with negligible charge asymmetry.  

\begin{figure}[tb]
\begin{center}
\begin{overpic}[width=0.49\linewidth]{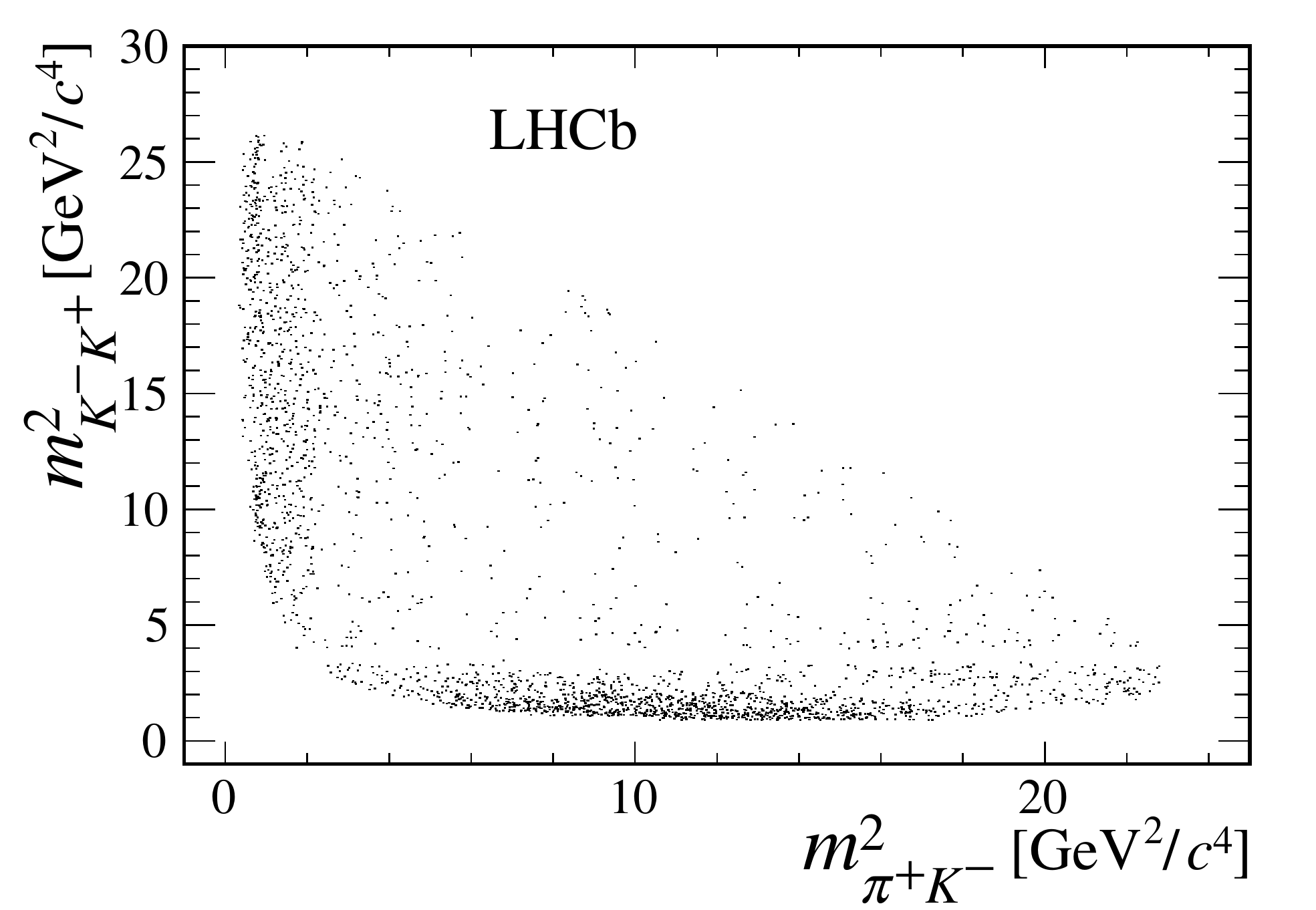}
\end{overpic}
\begin{overpic}[width=0.49\linewidth]{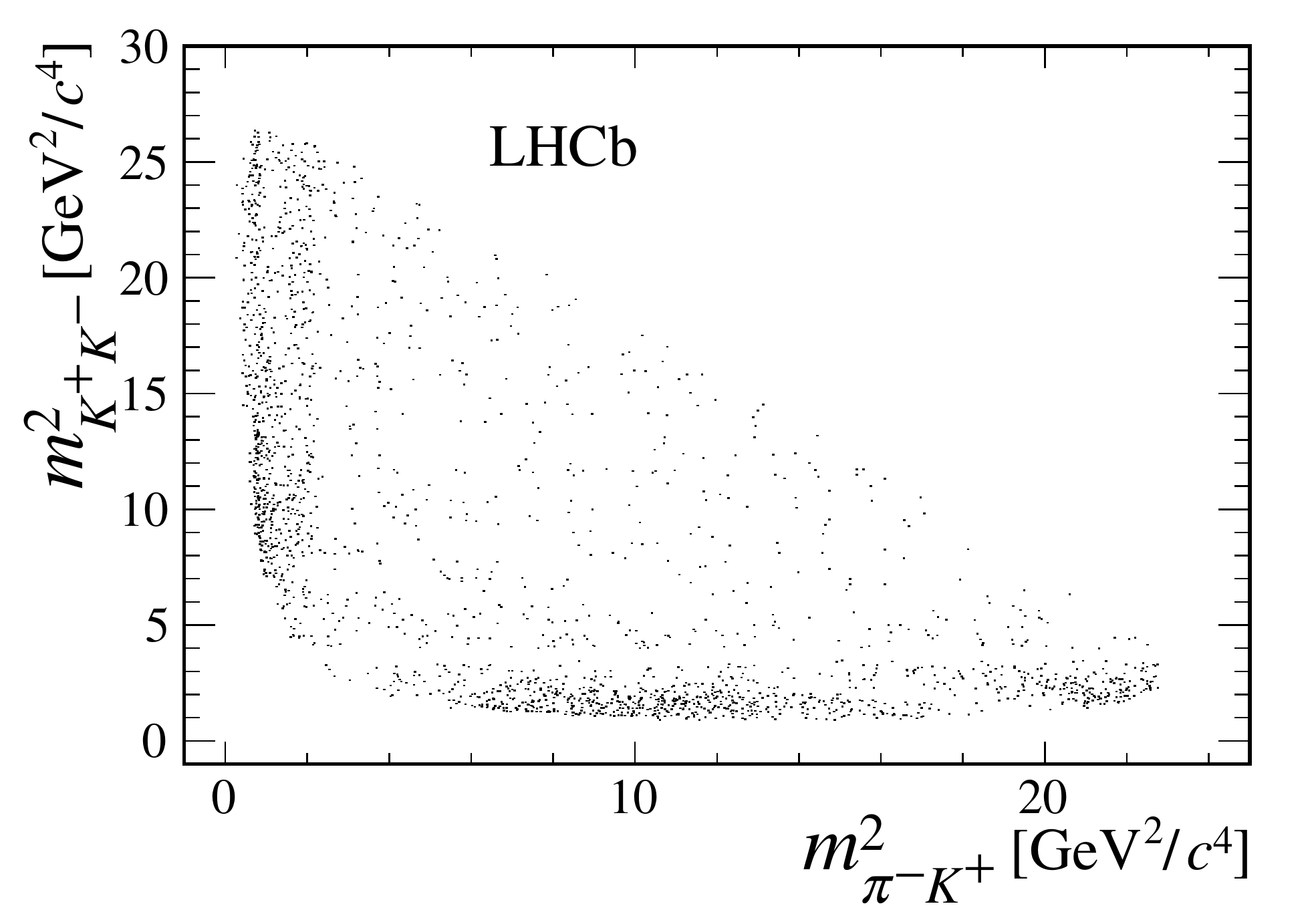}
\end{overpic}
 \end{center}
 \caption{
    {\small Dalitz plot for (left) \kkpip and (right) \kkpim candidates in the selected signal region.}
  }
  \label{fig:MyDPB}
\end{figure}

The distributions of the selected \Bpm  candidates, represented by the  Dalitz plot~\cite{Dalitz:1953cp}  constructed by the squared mass combinations \mmpik and \mmkk, are shown in Fig.~\ref{fig:MyDPB}. The clear differences between the \Bp and the \Bm distributions are due to \CPV effects~\cite{LHCb-PAPER-2014-044}.

%% file: models.tex

The total  \kkpip  decay amplitude, $\cal{A}$, can be expressed as function of $m^2_{\pi^+K^-}$  and \mmkk as
\begin{align}
{\mathcal{A}}(m^2_{\pi^+\!K^-}, \mmkk) = \sum^{N}_{i=1}{c_i}  {\mathcal{M}}_{i}(m^2_{\pi^+\!K^-}, \mmkk), 
\label{eq:totalAmplitude}
\end{align}
\noindent where $\mathcal{M}_{i}(m^2_{\pi^+\!K^-}, \mmkk)$ is the decay amplitude for an intermediate state~$i$. The analogous amplitude for the \Bm meson, $\overline{\cal{A}}$, is written in terms of $\overline{c}_i$ and  ${\overline{\mathcal{M}}}_{i}(m^2_{\pi^-\!K^+},\mmkk)$. 
This description for the total decay amplitude is known as the isobar model. 
In the amplitude fit, the complex coefficients  $c_i = (x_i + \Delta x_i) + i(y_i + \Delta y_i)$ and  $ \overline{c}_i = (x_i - \Delta x_i) + i(y_i - \Delta y_i)$ measure the relative contribution of each intermediate state $i$ for \Bp and \Bm, respectively, with $\Delta x_i$ and $\Delta y_i$ being the parameters that allow for  \CPV.   The individual amplitudes are described by
\begin{align}
\mathcal{M}_{i}(m^2_{\pi^+\!K^-}, \mmkk) = P_{i}(J, \vec{p}, \vec{q})F_B(|\vec{p}|)F_{i}(|\vec{q}|)T_{i}.
\label{eq:ampres}
\end{align}
The factor $P_{i}$ represents the angular part, which depends on the spin $J$ of the resonance. It is equal to 1, $-2\vec{p}\cdot \vec{q}$, and $\frac{4}{3}\left [ 3(\vec{p}\cdot \vec{q})^2 - (|\vec{p}| |\vec{q}|)^2 \right ]$,  for $J =0,1$ and $2$, respectively; $\vec{q}$ is the momentum of one of the resonance decay products and $\vec{p}$ is the momentum of the particle not forming the resonance,  both measured in the resonance rest frame. The Blatt--Weisskopf barrier factors~\cite{Blatt:1952ije,VonHippel:1972fg}, $F_B$ for the $B$ meson and $F_{i}$ for the resonance $i$,  account for  penetration effects due to the finite extent of the particles involved in the reaction. They are given by  1,  $ \sqrt{{(1 + z_0^2)}/{(1 + z^2)}}$ and $\sqrt{{(z_0^4 + 3z_0^2 + 9)}/{(z^4 + 3z^2 + 9)}}$ for $J =0,1$ and $2$, respectively, with $z = |\vec{q}| d$ or $z = |\vec{p}| d$  and $d$ the penetration radius, taken to be 4.0\,(GeV/$c$)$^{-1}$ $ \approx 0.8\fm$~\cite{LHCb-PAPER-2014-036,Aubert:2005ce}. The value of $z$ is $z_0$ when the invariant mass is equal to the nominal mass of the resonance. Finally, $T_{i}$ is a function representing the propagator of the intermediate state $i$. By default a relativistic Breit--Wigner function~\cite{Jackson:1964zd} is used,  which provides a good description for narrow resonances such as $K^{*}(892)^0$. More specific lineshapes are also used, as discussed further below. 

To determine the intermediate state contributions, a maximum-likelihood fit to the distribution of the \kkpi candidates in the Dalitz plot is performed using the {\sc Laura$^{++}$} package~\cite{Back:2017zqt,Ben-Haim:2014afa}. The total probability density function (PDF) is a sum of signal and background components, with relative contributions fixed from the result of the \kkpi mass fit. The background PDF is modelled according to its observed structures in the higher $m(\pipm\Kp\Km)$ sideband, the contribution from \kpipi crossfeed decays, using the model introduced by the \babar collaboration~\cite{Aubert:2008bj}, and an additional 0.6\% relative contribution from $\phi(1020)$ mesons randomly associated with a pion. The signal PDF for \Bp (\Bm) decays is given by $|{\cal A}|^{2}$ ($|\overline {\cal A}|^{2}$) multiplied by a function describing the variation of efficiency across the Dalitz plot. A histogram representing this efficiency map is obtained from simulated samples with corrections to account for known differences between data and simulation.
The $B^{+}$ and $B^{-}$ candidates are simultaneously fitted,  allowing for \CP violation. The \CP asymmetry, $A_{\CP_i}$, and fit fraction, ${\rm FF}_i$, for each component are given by
\begin{align}
  A_{\CP_i} = \frac{|\overline{c}_i|^2 - |c_i|^2}{|\overline{c}_i|^2 + |c_i|^2}  = \frac{-2(x_i\Delta x_i + y_i\Delta y_i)}{x_i^2 + (\Delta x_i)^2 + y_i^2 + (\Delta y_i)^2},
  \label{eq:acp}
\end{align}

\begin{align}
  {\rm FF}_i = \frac{\int ( |c_i \mathcal{M}_i|^{2}  + |\overline{c}_i\overline{\mathcal{M}}_i|^{2} )\text{d}\mmpik\text{d}\mmkk}{\int (|\mathcal{A}|^2 + |\mathcal{\overline{A}}|^2)\text{d}\mmpik\text{d}\mmkk}.
  \label{eq:FF}
\end{align}
The contribution of the possible intermediate states in the total decay amplitude is tested through a  procedure in which each component is taken in and out of the model, and that which provides the best likelihood is then maintained, and the process is repeated. In some regions of the phase space the observed signal yields could not be well described with only known resonance states and lineshapes, and thus alternative parameterisations were also tested.
 
In the $\pi^\pm K^\mp$ system, a nonresonant amplitude involving a single-pole form factor of the type (1+$m^2(\pi^\pm K^\mp)$/$\Lambda^2$)$^{-1}$, as proposed in ~\cite{Nogueira:2015tsa}, is included. This component, hereafter called single-pole amplitude, is a phenomenological description of the partonic interaction. 
The parameter $\Lambda$ sets the scale for the energy dependence and the proposed value of 1\gevcc  is used.

In the $K^+K^-$ system, a dedicated amplitude accounting for the $\pi \pi \leftrightarrow  KK$ rescattering is used. It is expressed as the product of the nonresonant single-pole form factor described above and a scattering term which accounts for the $S$-wave $\pi \pi \leftrightarrow KK$ transition amplitude, with isospin equal to 0 and $J=0$, given by the off-diagonal term in the S-matrix for the $\pi\pi$ and $KK$ coupled channel. The scattering term is expressed as $\sqrt{1-\nu^2}e^{2i\delta}$, where the inelasticity ($\nu$) and phase shift  ($\delta$) parametrisations are taken from Ref.~\cite{Pelaez:2004vs}. For the mass range 0.95 to 1.42\gevcc, where the coupling $\pi \pi \to KK$ is known to be important, these parameters are given by
 \begin{eqnarray}
\nu=1-\left(\epsilon_1\dfrac{k_2}{s^{1/2}}+\epsilon_2\dfrac{k_2^2}{s}\right)
\,\dfrac{M'^2-s}{s}
\end{eqnarray}

 and
\begin{eqnarray}
\cot\delta=C_0\,\dfrac{(s-M^2_s)(M^2_f-s)}{M^2_f s^{1/2}}\,
\dfrac{|k_2|}{k^2_2}, 
\end{eqnarray}
with parameters set as given in Ref.~\cite{Pelaez:2004vs}. 

  For all models tested in the analysis, the channel $B^\mp \to \KorKbar^*(892)^0 K^\mp$ is used as reference, with its real part $x$ fixed to one, $y$ and $\Delta y$ fixed to zero, while $\Delta x$ is free to vary. The values of  $x, y, \Delta x$ and $\Delta y$ for all other contributions are free parameters. The masses and widths of all resonances are fixed~\cite{PDG2018}.

%% file: results.tex

\begin{table*}[!hbt]
\caption{\small
\small Results of the Dalitz plot fit, where the first uncertainty is statistical and the second systematic. The fitted values of $c_i$ ($\bar c_i$) are expressed in terms of  magnitudes $|c_i|$ ($|\bar c_i|$) and phases $\arg(c_i)$ ($\arg(\bar c_i)$) for each \Bp (\Bm) contribution. The top row corresponds to \Bp and the bottom to \Bm mesons.
}
\label{tab:tab3013}
\resizebox{1.0\linewidth}{!}{

\input{tabs/tab_systematics_im.tex}

  }
\end{table*}

The fit results are summarised in Table~\ref{tab:tab3013}. Seven components are required to provide an overall good description of data; three of them correspond to the structure in the $\pi^\pm K^\mp$ system, and four for the \Kp \Km system. Statistical uncertainties are derived from the fitted values of $x, y, \Delta x, \Delta y$, with correlations and error propagation taken into account;
sources of systematic uncertainty are also evaluated as described later.

\begin{figure}[tb]
\begin{center}
\includegraphics[width=1.0\linewidth]{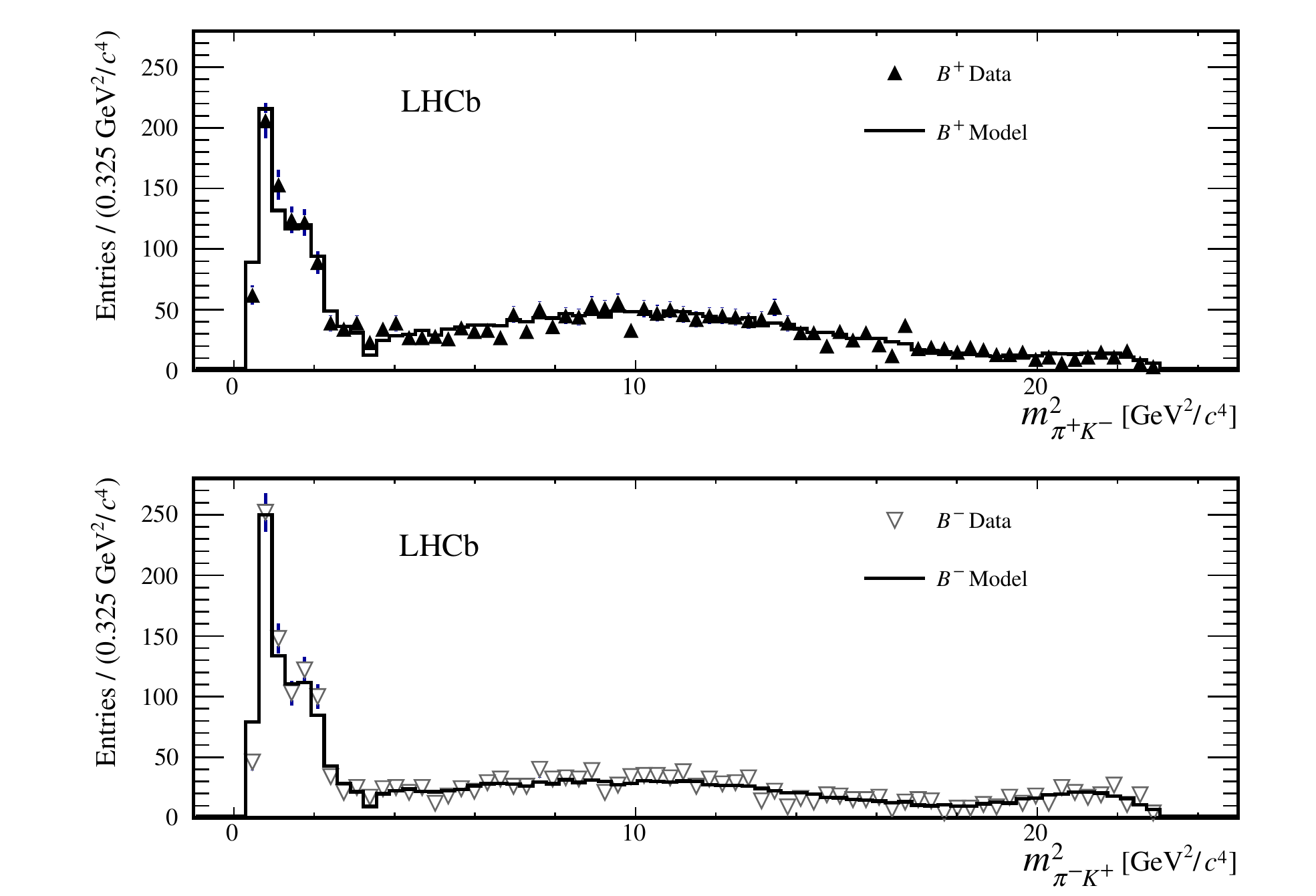}
 \end{center}
 \caption{
    \small Distribution of \mmpik. Data are represented by points for \Bp and \Bm candidates separately, with the result of the fit overlaid.
  }
  \label{fig:kpilow}
\end{figure}

The $\pi^\pm K^\mp$ system is well described by the contributions from the
$K^*(892)^0$ and $K^{*}_0(1430)^0$ resonances plus the single-pole  amplitude. The inclusion of the latter provides a better description of the data than that obtained from the $K^{*}_0(700)$, $K^{*}_2(1430)^0$, $K^{*}(1410)^0$,  and $K^{*}(1680)^0$ resonances. The largest contribution is from the single-pole amplitude with a total fit fraction of about 32\%. The   $K^{*}(892)^{0}$ and the $K^{*}_0(1430)^0$ amplitudes contribute to 7.5\% and 4.5\%, respectively. Given that they originate from penguin-diagram processes, their contributions to the total rate are expected to be small. The projection of the data onto \mmpik with the  fit model overlaid, is shown in Fig.~\ref{fig:kpilow}.

\begin{figure}[tb]
\begin{center}
\includegraphics[width=0.8\linewidth]{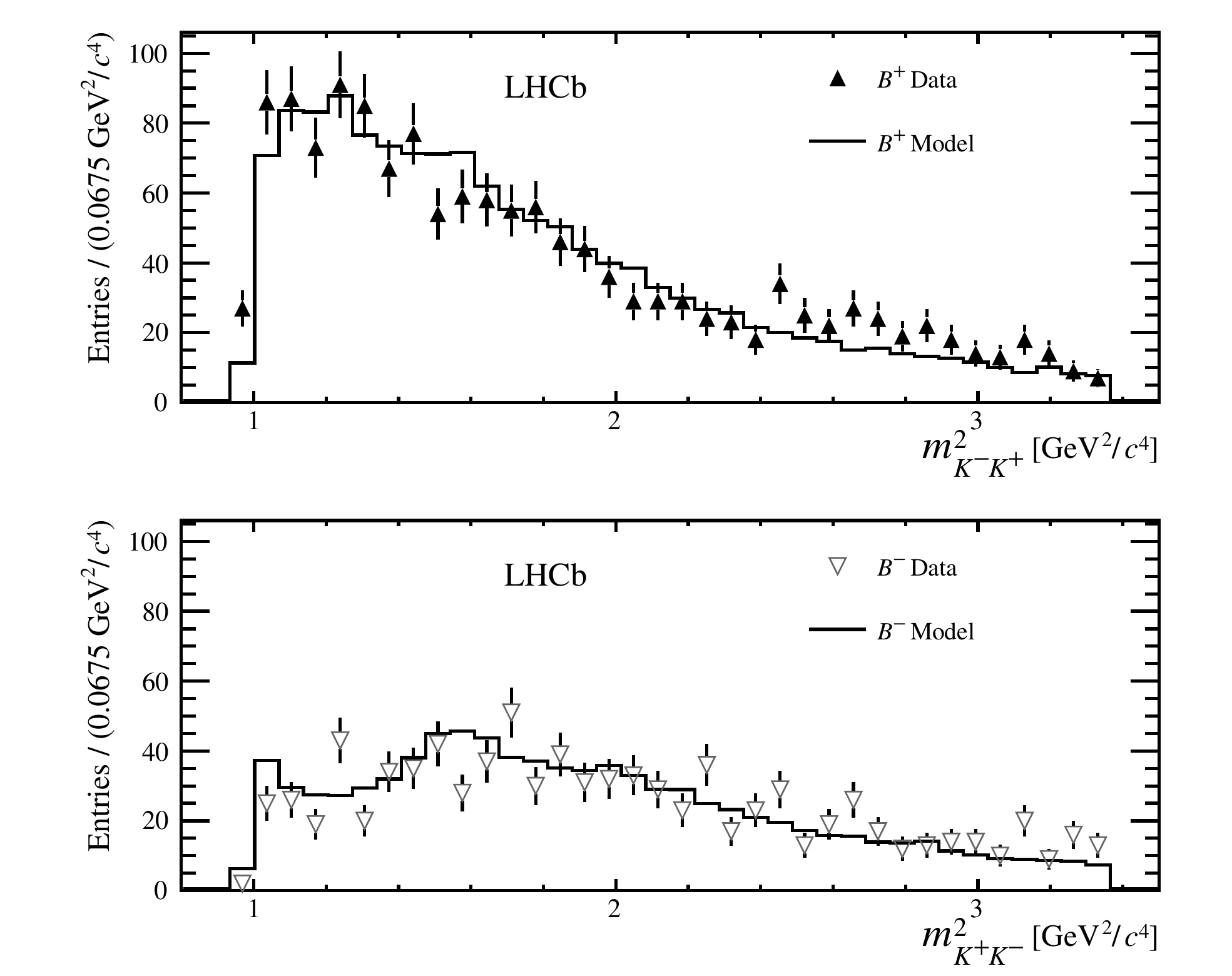}
 \end{center}
 \caption{
    \small Distribution of \mmkk up to 3.5\gevgevcccc. Data are represented by points for \Bp and \Bm separately, with the result of the fit overlaid.
  }
  \label{fig:kklow}
\end{figure}

In the $K^+K^-$ system, two main signatures can be highlighted: a strong destructive interference localised between $0.8$ and $3.3\gevgevcccc$ in \mmkk and projected between $12$ and $20\gevgevcccc$ in \mmpik, as shown in Fig.~\ref{fig:MyDPB}; and the large \CP asymmetry for \mmkk corresponding to the $\pi \pi \leftrightarrow KK$ rescattering region, as shown in Fig.~\ref{fig:kklow}. For the former, a good description of the data is achieved only when a high-mass vector amplitude is included in the Dalitz plot fit, producing the observed pattern through the interference with the $f_2(1270)$ amplitude. The data are well described by assuming this contribution to be the $\rho(1450)^0$ resonance, included in the fit with mass and width fixed to their known values~\cite{PDG2018}. The corresponding $B^\pm \to \rho(1450)^0\pi^\pm$ fit fraction is approximately 30\%, a rather large contribution not expected for the $K^+K^-$ pair as the dominant decay mode is $\pi\pi$ and the $\rho(1450)^0$ contribution in \pipipi is observed to be much lower~\cite{Aaij:2019jaq,Aaij:2019hzr}. A future analysis with the addition of the Run~2 data recorded with the \lhcb detector should be able to better pinpoint this effect.

With respect to the low \mmkk region, shown in Fig.~\ref{fig:kklow}, a significant contribution with a fit fraction of 16\% from the $\pi \pi \leftrightarrow KK$ $S$-wave rescattering amplitude is found. This contribution alone produces a  \CP asymmetry of $(-66 \pm 4 \pm 2)$\%, which is the largest \CPV manifestation ever observed for a single amplitude. This must be directly related to the total inclusive \CP asymmetry observed in this channel, which was previously reported to be $(-12.3 \pm 2.1)$\%. For the coupled channel \pipipi, with a branching fraction three times larger than that of \kkpi, a positive \CP asymmetry has been measured~\cite{LHCb-PAPER-2014-044}. This gives consistency for  the interpretation of the large \CPV observed here originates from rescattering effects. 
Finally,  the inclusion of the $\phi(1020)$ resonance in the amplitude model also improves the data description near the \Kp\Km threshold, however with a statistically marginal contribution. The model is also not perfect in other regions in \mmkk, for instance for \Bp decays in a few bins above 2.5\gevgevcccc.

A second solution is found in the fit, presenting a large \CP asymmetry of 76\% in the $K^{*}_0(1430)^0$ component, compensated by a similarly large negative asymmetry in the interference term between the $K^{*}_0(1430)^0$ and the single-pole amplitudes. The net effect is a negligible \CP asymmetry near the $K^{*}_0(1430)^0$ region, matching what is seen data. This solution presents a large sum of fit fractions for the \Bm decay, of about 130\%,  indicating this is probably a fake effect created by the fit. As such, this solution is interpreted as unphysical. More data are necessary to understand this feature.

Several sources of systematic uncertainty are considered. These include possible mismodellings in the mass fit,  the efficiency variation and background description across the Dalitz plot, the uncertainty associated to the fixed parameters in the Dalitz plot fit and possible biases in the fitting procedure.

The impact of the systematic studies affect differently each of the amplitudes. The main contribution comes from the variation of the masses and widths of the resonances; their central values and uncertainties are taken from Ref.~\cite{PDG2018} and are randomised according to a Gaussian distribution. This effect is particularly important for the $K^{*}_0(1430)^0$ and single-pole components, the two broad scalar contributions in the $\pi^\pm K^\mp$ system. The absolute uncertainties on their fractions are found to be 0.8\% and 3.0\%, respectively. The second main contribution comes from the $\pi^{\pm}K^+K^-$ mass fit, impacting most on the $K^{*}(892)^0$, $K^{*}_0(1430)^0$ and single-pole fractions with uncertainties of 0.4\%, 0.8\% and 2.0\%, respectively. The systematic uncertainty associated to efficiency variation across the Dalitz plot is studied by performing several fits to data with efficiency maps obtained by varying the bin contents of the original efficiency histogram according to their uncertainty; this results in uncertainties in the fit fractions that range from 0.01\% to 0.1\%. The systematic uncertainty due to the background models is evaluated with a similar procedure, also resulting in small uncertainties. The \Bpm production and kaon detection asymmetry effects are taken into account following Ref.~\cite{LHCb-PAPER-2014-013}, with associated uncertainties less than 0.1\%.  All systematic uncertainties are added in quadrature and represent the second uncertainty in Table~\ref{tab:tab3013}.

%% file: tabs/tab_systematics_im.tex
\begin{tabular}
{@{\hspace{0.5cm}}c@{\hspace{0.25cm}}  @{\hspace{0.25cm}}c@{\hspace{0.25cm}}  @{\hspace{0.25cm}}c@{\hspace{0.25cm}}  @{\hspace{0.25cm}}c@{\hspace{0.25cm}}  @{\hspace{0.25cm}}l@{\hspace{0.25cm}}  @{\hspace{0.25cm}}l@{\hspace{0.5cm}}}
      \hline \hline
      Contribution    & Fit Fraction(\%)        & $A_{\CP}$(\%)               & Magnitude (\Bp/\Bm)                                         & Phase [$\degrees$] (\Bp/\Bm)                        \\ \hline
       $K^*(892)^0$     & $\phantom{0}7.5 \pm 0.6 \pm 0.5$  & $+12.3\pm \phantom{1}8.7 \pm \phantom{0}4.5$     & $0.94 \pm 0.04 \pm 0.02$   & $\phantom{-00}0$ (fixed)    \\
                                    &                                                          &                                                                  &      $1.06 \pm 0.04 \pm 0.02$ & $\phantom{-00}0$ (fixed) \\ \hline
      $K^{*}_0(1430)^0$   & $\phantom{0}4.5 \pm 0.7\pm 1.2$   & $+10.4 \pm 14.9  \pm \phantom{0}8.8$    & $0.74 \pm 0.09 \pm 0.09$ &  $-176  \pm 10 \pm 16$ \\
                                   &                                                          &                                                                  & $0.82 \pm 0.09 \pm 0.10$  & $\phantom{-}136 \pm 11 \pm 21$     \\ \hline
      
       Single pole  &  $32.3 \pm 1.5 \pm 4.1$   & $-10.7 \pm \phantom{0}5.3 \pm \phantom{0}3.5$    & $2.19 \pm 0.13 \pm 0.17 $ & $-138 \pm \phantom{0}7 \pm \phantom{0}5$  \\
                           &                                          &                                                                                  &  $ 1.97 \pm 0.12 \pm 0.20$  & $\phantom{-}166 \pm \phantom{0}6 \pm \phantom{0}5$ \\ \hline
      $\rho(1450)^0$    & $30.7 \pm 1.2 \pm 0.9$   & $-10.9 \pm \phantom{0}4.4 \pm \phantom{0}2.4$  & $2.14 \pm 0.11 \pm 0.07$ & $-175 \pm 10 \pm 15$ \\
                           &                                          &                                                                                  & $1.92 \pm 0.10 \pm 0.07$ &  $\phantom{-}140 \pm 13 \pm 20$ \\ \hline
      $f_2(1270)$  & $\phantom{0}7.5 \pm 0.8 \pm 0.7$   & $+26.7 \pm 10.2 \pm \phantom{0}4.8$ & $0.86 \pm 0.09 \pm 0.07 $& $-106 \pm 11 \pm 10$  \\
                           &                                          &                                                                                  & $1.13 \pm 0.08 \pm 0.05$   & $-128 \pm 11 \pm 14$ \\ \hline
      $\rm{Rescattering}$  & $16.4 \pm 0.8 \pm 1.0$   & $-66.4 \pm \phantom{0}3.8 \pm \phantom{0}1.9$ & $1.91 \pm 0.09 \pm 0.06$ & \phantom{-}$\,-56 \pm 12 \pm 18$ \\
                           &                                          &                                                                                  & $0.86 \pm 0.07 \pm 0.04$ & \phantom{0}$-81 \pm 14 \pm 15$ \\ \hline
       $\phi(1020)$  &\phantom{0}$0.3 \pm 0.1 \pm 0.1$   & $\,\;+9.8 \pm 43.6 \pm 26.6$ & $0.20 \pm 0.07 \pm 0.02 $  & \phantom{0}$-52 \pm 23 \pm 32$ \\ 
                          &                                          &                                                                                  & $0.22\pm 0.06 \pm 0.04$ & $\phantom{-}107 \pm 33 \pm 41$ \\
      \hline \hline
\end{tabular}

%% file: conclusions.tex

In summary, the resonant substructure of the charmless three-body \kkpi decay is determined using the isobar model formalism, providing an overall good description of the observed data. Three components are obtained for the $\pi^\pm K^\mp$ system: two resonant states ($K^*(892)^0$, $K^{*}_0(1430)^0$) with a \CP asymmetry consistent with zero, and a nonresonant single-pole form factor contribution with a fit fraction of about $30$\%. Two other components are found, $\rho(1450)$ and $f_2(1270)$, which provide a destructive interference pattern in the Dalitz plot. The rescattering amplitude, acting in the region $0.95 < \mkk < 1.42$\,\gevcc, produces a negative \CP asymmetry of $(-66 \pm 4 \pm 2)\%$, which is the largest \CP violation effect observed from a single amplitude.

%% file: acknowledgements.tex
\section*{Acknowledgements}
%
%
\noindent We express our gratitude to our colleagues in the CERN
accelerator departments for the excellent performance of the LHC. We
thank the technical and administrative staff at the LHCb
institutes.
We acknowledge support from CERN and from the national agencies:
CAPES, CNPq, FAPERJ and FINEP (Brazil); 
MOST and NSFC (China); 
CNRS/IN2P3 (France); 
BMBF, DFG and MPG (Germany); 
INFN (Italy); 
NWO (Netherlands); 
MNiSW and NCN (Poland); 
MEN/IFA (Romania); 
MSHE (Russia); 
MinECo (Spain); 
SNSF and SER (Switzerland); 
NASU (Ukraine); 
STFC (United Kingdom); 
DOE NP and NSF (USA).
We acknowledge the computing resources that are provided by CERN, IN2P3
(France), KIT and DESY (Germany), INFN (Italy), SURF (Netherlands),
PIC (Spain), GridPP (United Kingdom), RRCKI and Yandex
LLC (Russia), CSCS (Switzerland), IFIN-HH (Romania), CBPF (Brazil),
PL-GRID (Poland) and OSC (USA).
We are indebted to the communities behind the multiple open-source
software packages on which we depend.
Individual groups or members have received support from
AvH Foundation (Germany);
EPLANET, Marie Sk\l{}odowska-Curie Actions and ERC (European Union);
ANR, Labex P2IO and OCEVU, and R\'{e}gion Auvergne-Rh\^{o}ne-Alpes (France);
Key Research Program of Frontier Sciences of CAS, CAS PIFI, and the Thousand Talents Program (China);
RFBR, RSF and Yandex LLC (Russia);
GVA, XuntaGal and GENCAT (Spain);
the Royal Society
and the Leverhulme Trust (United Kingdom).

%% file: LHCb_Authorship_13-Dec-2018.tex
\centerline
{\large\bf LHCb Collaboration}
\begin
{flushleft}
\small
R.~Aaij$^{29}$,
C.~Abell{\'a}n~Beteta$^{46}$,
B.~Adeva$^{43}$,
M.~Adinolfi$^{50}$,
C.A.~Aidala$^{77}$,
Z.~Ajaltouni$^{7}$,
S.~Akar$^{61}$,
P.~Albicocco$^{20}$,
J.~Albrecht$^{12}$,
F.~Alessio$^{44}$
% ...

{\footnotesize \it

% ...
$ ^{7}$Universit{\'e} Clermont Auvergne, CNRS/IN2P3, LPC, Clermont-Ferrand, France\\
% ...
$ ^{12}$Fakult{\"a}t Physik, Technische Universit{\"a}t Dortmund, Dortmund, Germany\\
% ...
$ ^{20}$INFN Laboratori Nazionali di Frascati, Frascati, Italy\\
% ...
$ ^{29}$Nikhef National Institute for Subatomic Physics, Amsterdam, Netherlands\\
% ...
$ ^{43}$Instituto Galego de F{\'\i}sica de Altas Enerx{\'\i}as (IGFAE), Universidade de Santiago de Compostela, Santiago de Compostela, Spain\\
$ ^{44}$European Organization for Nuclear Research (CERN), Geneva, Switzerland\\
% ...
$ ^{46}$Physik-Institut, Universit{\"a}t Z{\"u}rich, Z{\"u}rich, Switzerland\\
% ...
$ ^{50}$H.H. Wills Physics Laboratory, University of Bristol, Bristol, United Kingdom\\
% ...
$ ^{61}$University of Cincinnati, Cincinnati, OH, United States\\
% ...
$ ^{63}$Syracuse University, Syracuse, NY, United States\\
% ...
$ ^{77}$University of Michigan, Ann Arbor, United States, associated to $^{63}$\\
% ...
}
\end{flushleft}

%% file: main.bbl
\ifx\mcitethebibliography\mciteundefinedmacro
\PackageError{LHCb.bst}{mciteplus.sty has not been loaded}
{This bibstyle requires the use of the mciteplus package.}\fi
\providecommand{\href}[2]{#2}
\begin{mcitethebibliography}{10}
\mciteSetBstSublistMode{n}
\mciteSetBstMaxWidthForm{subitem}{\alph{mcitesubitemcount})}
\mciteSetBstSublistLabelBeginEnd{\mcitemaxwidthsubitemform\space}
{\relax}{\relax}

\bibitem{LHCb-PAPER-2014-044}
LHCb collaboration, R.~Aaij {\em et~al.},
  \ifthenelse{\boolean{articletitles}}{\emph{{Measurement of \CP violation in
  the three-body phase space of charmless \Bpm decays}},
  }{}\href{https://doi.org/10.1103/PhysRevD.90.112004}{Phys.\ Rev.\
  \textbf{D90} (2014) 112004},
  \href{http://arxiv.org/abs/1408.5373}{{\normalfont\ttfamily
  arXiv:1408.5373}}\relax
\mciteBstWouldAddEndPuncttrue
\mciteSetBstMidEndSepPunct{\mcitedefaultmidpunct}
{\mcitedefaultendpunct}{\mcitedefaultseppunct}\relax
\EndOfBibitem
\bibitem{Hsu:2017kir}
Belle collaboration, C.~L. Hsu {\em et~al.},
  \ifthenelse{\boolean{articletitles}}{\emph{{Measurement of branching fraction
  and direct \CP asymmetry in charmless $B^+ \to K^+K^- \pi^+$ decays at
  Belle}}, }{}\href{https://doi.org/10.1103/PhysRevD.96.031101}{Phys.\ Rev.\
  \textbf{D96} (2017) 031101},
  \href{http://arxiv.org/abs/1705.02640}{{\normalfont\ttfamily
  arXiv:1705.02640}}\relax
\mciteBstWouldAddEndPuncttrue
\mciteSetBstMidEndSepPunct{\mcitedefaultmidpunct}
{\mcitedefaultendpunct}{\mcitedefaultseppunct}\relax
\EndOfBibitem
\bibitem{Aubert:2007xb}
BaBar collaboration, B.~Aubert {\em et~al.},
  \ifthenelse{\boolean{articletitles}}{\emph{{Observation of the Decay $B^{+}
  \to K^{+} K^{-} \pi^{+}$}},
  }{}\href{https://doi.org/10.1103/PhysRevLett.99.221801}{Phys.\ Rev.\ Lett.\
  \textbf{99} (2007) 221801},
  \href{http://arxiv.org/abs/0708.0376}{{\normalfont\ttfamily
  arXiv:0708.0376}}\relax
\mciteBstWouldAddEndPuncttrue
\mciteSetBstMidEndSepPunct{\mcitedefaultmidpunct}
{\mcitedefaultendpunct}{\mcitedefaultseppunct}\relax
\EndOfBibitem
\bibitem{Aubert:2009av}
BaBar collaboration, B.~Aubert {\em et~al.},
  \ifthenelse{\boolean{articletitles}}{\emph{{Dalitz Plot Analysis of $B^{\pm}
  \to \pi^{\pm}\pi^{\pm}\pi^{\mp}$ Decays}},
  }{}\href{https://doi.org/10.1103/PhysRevD.79.072006}{Phys.\ Rev.\
  \textbf{D79} (2009) 072006},
  \href{http://arxiv.org/abs/0902.2051}{{\normalfont\ttfamily
  arXiv:0902.2051}}\relax
\mciteBstWouldAddEndPuncttrue
\mciteSetBstMidEndSepPunct{\mcitedefaultmidpunct}
{\mcitedefaultendpunct}{\mcitedefaultseppunct}\relax
\EndOfBibitem
\bibitem{Lees:2012kxa}
BaBar collaboration, J.~P. Lees {\em et~al.},
  \ifthenelse{\boolean{articletitles}}{\emph{{Study of \CP violation in
  Dalitz-plot analyses of $B^0 \to K^+K^-K^0_S, B^+ \to K^+K^-K^+$, and $B^+
  \to K^0_SK^0_SK^+$}},
  }{}\href{https://doi.org/10.1103/PhysRevD.85.112010}{Phys.\ Rev.\
  \textbf{D85} (2012) 112010},
  \href{http://arxiv.org/abs/1201.5897}{{\normalfont\ttfamily
  arXiv:1201.5897}}\relax
\mciteBstWouldAddEndPuncttrue
\mciteSetBstMidEndSepPunct{\mcitedefaultmidpunct}
{\mcitedefaultendpunct}{\mcitedefaultseppunct}\relax
\EndOfBibitem
\bibitem{Aubert:2008bj}
BaBar collaboration, B.~Aubert {\em et~al.},
  \ifthenelse{\boolean{articletitles}}{\emph{{Evidence for direct \CP violation
  from Dalitz-plot analysis of $B^\pm \to K^\pm \pi^\mp \pi^\pm$}},
  }{}\href{https://doi.org/10.1103/PhysRevD.78.012004}{Phys.\ Rev.\
  \textbf{D78} (2008) 012004},
  \href{http://arxiv.org/abs/0803.4451}{{\normalfont\ttfamily
  arXiv:0803.4451}}\relax
\mciteBstWouldAddEndPuncttrue
\mciteSetBstMidEndSepPunct{\mcitedefaultmidpunct}
{\mcitedefaultendpunct}{\mcitedefaultseppunct}\relax
\EndOfBibitem
\bibitem{Garmash:2005rv}
Belle collaboration, A.~Garmash {\em et~al.},
  \ifthenelse{\boolean{articletitles}}{\emph{{Evidence for large direct CP
  violation in $B^{\pm} \to \rho(770)^0K^{\pm}$ from analysis of the three-body
  charmless $B^{\pm} \to K^{\pm} \pi^{\pm} \pi^{\mp}$ decay}},
  }{}\href{https://doi.org/10.1103/PhysRevLett.96.251803}{Phys.\ Rev.\ Lett.\
  \textbf{96} (2006) 251803},
  \href{http://arxiv.org/abs/hep-ex/0512066}{{\normalfont\ttfamily
  arXiv:hep-ex/0512066}}\relax
\mciteBstWouldAddEndPuncttrue
\mciteSetBstMidEndSepPunct{\mcitedefaultmidpunct}
{\mcitedefaultendpunct}{\mcitedefaultseppunct}\relax
\EndOfBibitem
\bibitem{Okubo:1963fa}
S.~Okubo, \ifthenelse{\boolean{articletitles}}{\emph{{$\phi$-meson and unitary
  symmetry model}},
  }{}\href{https://doi.org/10.1016/S0375-9601(63)92548-9}{Phys.\ Lett.\
  \textbf{5} (1963) 165}\relax
\mciteBstWouldAddEndPuncttrue
\mciteSetBstMidEndSepPunct{\mcitedefaultmidpunct}
{\mcitedefaultendpunct}{\mcitedefaultseppunct}\relax
\EndOfBibitem
\bibitem{Zweig:352337}
G.~Zweig, \ifthenelse{\boolean{articletitles}}{\emph{{An SU$_3$ model for
  strong interaction symmetry and its breaking; Version 1}}, }{} Tech. Rep.
  \href{http://cds.cern.ch/record/352337}{CERN-TH-401}, CERN, Geneva,
  1964\relax
\mciteBstWouldAddEndPuncttrue
\mciteSetBstMidEndSepPunct{\mcitedefaultmidpunct}
{\mcitedefaultendpunct}{\mcitedefaultseppunct}\relax
\EndOfBibitem
\bibitem{Iizuka:1966fk}
J.~Iizuka, \ifthenelse{\boolean{articletitles}}{\emph{{Systematics and
  phenomenology of meson family}},
  }{}\href{https://doi.org/10.1143/PTPS.37.21}{Prog.\ Theor.\ Phys.\ Suppl.\
  \textbf{37} (1966) 21}\relax
\mciteBstWouldAddEndPuncttrue
\mciteSetBstMidEndSepPunct{\mcitedefaultmidpunct}
{\mcitedefaultendpunct}{\mcitedefaultseppunct}\relax
\EndOfBibitem
\bibitem{LHCb-PAPER-2013-048}
LHCb collaboration, R.~Aaij {\em et~al.},
  \ifthenelse{\boolean{articletitles}}{\emph{{Measurement of the charge
  asymmetry in \mbox{\decay{\Bpm}{\phiz\Kpm}} and search for
  \mbox{\decay{\Bpm}{\phiz\pipm}} decays}},
  }{}\href{https://doi.org/10.1016/j.physletb.2013.11.036}{Phys.\ Lett.\
  \textbf{B728} (2014) 85},
  \href{http://arxiv.org/abs/1309.3742}{{\normalfont\ttfamily
  arXiv:1309.3742}}\relax
\mciteBstWouldAddEndPuncttrue
\mciteSetBstMidEndSepPunct{\mcitedefaultmidpunct}
{\mcitedefaultendpunct}{\mcitedefaultseppunct}\relax
\EndOfBibitem
\bibitem{Estabrooks:1973zd}
P.~Estabrooks {\em et~al.}, \ifthenelse{\boolean{articletitles}}{\emph{{$\pi
  \pi$ phase shift analysis from 600 to 1900 MeV}},
  }{}\href{https://doi.org/10.1063/1.2947103}{AIP Conf.\ Proc.\  \textbf{13}
  (1973) 206}\relax
\mciteBstWouldAddEndPuncttrue
\mciteSetBstMidEndSepPunct{\mcitedefaultmidpunct}
{\mcitedefaultendpunct}{\mcitedefaultseppunct}\relax
\EndOfBibitem
\bibitem{Cohen:1980cq}
D.~H. Cohen {\em et~al.}, \ifthenelse{\boolean{articletitles}}{\emph{{Amplitude
  Analysis of the \Km \Kp system produced in the reactions \pim $p$ $\to$ \Km
  \Kp $n$ and \pip $n$ $\to$ \Km \Kp $p$ at 6-\gevc}},
  }{}\href{https://doi.org/10.1103/PhysRevD.22.2595}{Phys.\ Rev.\  \textbf{D22}
  (1980) 2595}\relax
\mciteBstWouldAddEndPuncttrue
\mciteSetBstMidEndSepPunct{\mcitedefaultmidpunct}
{\mcitedefaultendpunct}{\mcitedefaultseppunct}\relax
\EndOfBibitem
\bibitem{Nogueira:2015tsa}
J.~H. Alvarenga~Nogueira {\em et~al.},
  \ifthenelse{\boolean{articletitles}}{\emph{{\CP violation: Dalitz
  interference, $CPT$, and final state interactions}},
  }{}\href{https://doi.org/10.1103/PhysRevD.92.054010}{Phys.\ Rev.\
  \textbf{D92} (2015) 054010}\relax
\mciteBstWouldAddEndPuncttrue
\mciteSetBstMidEndSepPunct{\mcitedefaultmidpunct}
{\mcitedefaultendpunct}{\mcitedefaultseppunct}\relax
\EndOfBibitem
\bibitem{Wolfenstein:1990ks}
L.~Wolfenstein, \ifthenelse{\boolean{articletitles}}{\emph{{Final state
  interactions and CP violation in weak decays}},
  }{}\href{https://doi.org/10.1103/PhysRevD.43.151}{Phys.\ Rev.\  \textbf{D43}
  (1991) 151}\relax
\mciteBstWouldAddEndPuncttrue
\mciteSetBstMidEndSepPunct{\mcitedefaultmidpunct}
{\mcitedefaultendpunct}{\mcitedefaultseppunct}\relax
\EndOfBibitem
\bibitem{Bigi:2000yz}
I.~I. Bigi and A.~I. Sanda, \ifthenelse{\boolean{articletitles}}{\emph{{\CP
  violation}}, }{}\href{https://doi.org/10.1017/CBO9780511581014}{Camb.\
  Monogr.\ Part.\ Phys.\ Nucl.\ Phys.\ Cosmol.\ \ \textbf{9} (2009) 1}\relax
\mciteBstWouldAddEndPuncttrue
\mciteSetBstMidEndSepPunct{\mcitedefaultmidpunct}
{\mcitedefaultendpunct}{\mcitedefaultseppunct}\relax
\EndOfBibitem
\bibitem{Fleming:1964zz}
G.~N. Fleming, \ifthenelse{\boolean{articletitles}}{\emph{{Recoupling Effects
  in the Isobar Model. 1. General Formalism for Three-Pion Scattering}},
  }{}\href{https://doi.org/10.1103/PhysRev.135.B551}{Phys.\ Rev.\  \textbf{135}
  (1964) B551}\relax
\mciteBstWouldAddEndPuncttrue
\mciteSetBstMidEndSepPunct{\mcitedefaultmidpunct}
{\mcitedefaultendpunct}{\mcitedefaultseppunct}\relax
\EndOfBibitem
\bibitem{Herndon:1973yn}
D.~Herndon, P.~Soding, and R.~J. Cashmore,
  \ifthenelse{\boolean{articletitles}}{\emph{{Generalized isobar model
  formalism}}, }{}\href{https://doi.org/10.1103/PhysRevD.11.3165}{Phys.\ Rev.\
  \textbf{D11} (1975) 3165}\relax
\mciteBstWouldAddEndPuncttrue
\mciteSetBstMidEndSepPunct{\mcitedefaultmidpunct}
{\mcitedefaultendpunct}{\mcitedefaultseppunct}\relax
\EndOfBibitem
\bibitem{LHCb-DP-2008-001}
LHCb collaboration, A.~A. Alves~Jr.\ {\em et~al.},
  \ifthenelse{\boolean{articletitles}}{\emph{{The \lhcb detector at the LHC}},
  }{}\href{https://doi.org/10.1088/1748-0221/3/08/S08005}{JINST \textbf{3}
  (2008) S08005}\relax
\mciteBstWouldAddEndPuncttrue
\mciteSetBstMidEndSepPunct{\mcitedefaultmidpunct}
{\mcitedefaultendpunct}{\mcitedefaultseppunct}\relax
\EndOfBibitem
\bibitem{LHCb-DP-2014-002}
LHCb collaboration, R.~Aaij {\em et~al.},
  \ifthenelse{\boolean{articletitles}}{\emph{{LHCb detector performance}},
  }{}\href{https://doi.org/10.1142/S0217751X15300227}{Int.\ J.\ Mod.\ Phys.\
  \textbf{A30} (2015) 1530022},
  \href{http://arxiv.org/abs/1412.6352}{{\normalfont\ttfamily
  arXiv:1412.6352}}\relax
\mciteBstWouldAddEndPuncttrue
\mciteSetBstMidEndSepPunct{\mcitedefaultmidpunct}
{\mcitedefaultendpunct}{\mcitedefaultseppunct}\relax
\EndOfBibitem
\bibitem{Sjostrand:2007gs}
T.~Sj\"{o}strand, S.~Mrenna, and P.~Skands,
  \ifthenelse{\boolean{articletitles}}{\emph{{A brief introduction to PYTHIA
  8.1}}, }{}\href{https://doi.org/10.1016/j.cpc.2008.01.036}{Comput.\ Phys.\
  Commun.\  \textbf{178} (2008) 852},
  \href{http://arxiv.org/abs/0710.3820}{{\normalfont\ttfamily
  arXiv:0710.3820}}\relax
\mciteBstWouldAddEndPuncttrue
\mciteSetBstMidEndSepPunct{\mcitedefaultmidpunct}
{\mcitedefaultendpunct}{\mcitedefaultseppunct}\relax
\EndOfBibitem
\bibitem{LHCb-PROC-2010-056}
I.~Belyaev {\em et~al.}, \ifthenelse{\boolean{articletitles}}{\emph{{Handling
  of the generation of primary events in Gauss, the LHCb simulation
  framework}}, }{}\href{https://doi.org/10.1088/1742-6596/331/3/032047}{J.\
  Phys.\ Conf.\ Ser.\  \textbf{331} (2011) 032047}\relax
\mciteBstWouldAddEndPuncttrue
\mciteSetBstMidEndSepPunct{\mcitedefaultmidpunct}
{\mcitedefaultendpunct}{\mcitedefaultseppunct}\relax
\EndOfBibitem
\bibitem{Lange:2001uf}
D.~J. Lange, \ifthenelse{\boolean{articletitles}}{\emph{{The EvtGen particle
  decay simulation package}},
  }{}\href{https://doi.org/10.1016/S0168-9002(01)00089-4}{Nucl.\ Instrum.\
  Meth.\  \textbf{A462} (2001) 152}\relax
\mciteBstWouldAddEndPuncttrue
\mciteSetBstMidEndSepPunct{\mcitedefaultmidpunct}
{\mcitedefaultendpunct}{\mcitedefaultseppunct}\relax
\EndOfBibitem
\bibitem{Golonka:2005pn}
P.~Golonka and Z.~Was, \ifthenelse{\boolean{articletitles}}{\emph{{PHOTOS Monte
  Carlo: A precision tool for QED corrections in $Z$ and $W$ decays}},
  }{}\href{https://doi.org/10.1140/epjc/s2005-02396-4}{Eur.\ Phys.\ J.\
  \textbf{C45} (2006) 97},
  \href{http://arxiv.org/abs/hep-ph/0506026}{{\normalfont\ttfamily
  arXiv:hep-ph/0506026}}\relax
\mciteBstWouldAddEndPuncttrue
\mciteSetBstMidEndSepPunct{\mcitedefaultmidpunct}
{\mcitedefaultendpunct}{\mcitedefaultseppunct}\relax
\EndOfBibitem
\bibitem{Allison:2006ve}
Geant4 collaboration, J.~Allison {\em et~al.},
  \ifthenelse{\boolean{articletitles}}{\emph{{Geant4 developments and
  applications}}, }{}\href{https://doi.org/10.1109/TNS.2006.869826}{IEEE
  Trans.\ Nucl.\ Sci.\  \textbf{53} (2006) 270}\relax
\mciteBstWouldAddEndPuncttrue
\mciteSetBstMidEndSepPunct{\mcitedefaultmidpunct}
{\mcitedefaultendpunct}{\mcitedefaultseppunct}\relax
\EndOfBibitem
\bibitem{Agostinelli:2002hh}
Geant4 collaboration, S.~Agostinelli {\em et~al.},
  \ifthenelse{\boolean{articletitles}}{\emph{{Geant4: A simulation toolkit}},
  }{}\href{https://doi.org/10.1016/S0168-9002(03)01368-8}{Nucl.\ Instrum.\
  Meth.\  \textbf{A506} (2003) 250}\relax
\mciteBstWouldAddEndPuncttrue
\mciteSetBstMidEndSepPunct{\mcitedefaultmidpunct}
{\mcitedefaultendpunct}{\mcitedefaultseppunct}\relax
\EndOfBibitem
\bibitem{LHCb-PROC-2011-006}
M.~Clemencic {\em et~al.}, \ifthenelse{\boolean{articletitles}}{\emph{{The
  \lhcb simulation application, Gauss: Design, evolution and experience}},
  }{}\href{https://doi.org/10.1088/1742-6596/331/3/032023}{J.\ Phys.\ Conf.\
  Ser.\  \textbf{331} (2011) 032023}\relax
\mciteBstWouldAddEndPuncttrue
\mciteSetBstMidEndSepPunct{\mcitedefaultmidpunct}
{\mcitedefaultendpunct}{\mcitedefaultseppunct}\relax
\EndOfBibitem
\bibitem{PDG2018}
Particle Data Group, M.~Tanabashi {\em et~al.},
  \ifthenelse{\boolean{articletitles}}{\emph{{\href{http://pdg.lbl.gov/}{Review
  of particle physics}}},
  }{}\href{https://doi.org/10.1103/PhysRevD.98.030001}{Phys.\ Rev.\
  \textbf{D98} (2018) 030001}\relax
\mciteBstWouldAddEndPuncttrue
\mciteSetBstMidEndSepPunct{\mcitedefaultmidpunct}
{\mcitedefaultendpunct}{\mcitedefaultseppunct}\relax
\EndOfBibitem
\bibitem{Breiman}
L.~Breiman, J.~H. Friedman, R.~A. Olshen, and C.~J. Stone, {\em Classification
  and regression trees}, Wadsworth international group, Belmont, California,
  USA, 1984\relax
\mciteBstWouldAddEndPuncttrue
\mciteSetBstMidEndSepPunct{\mcitedefaultmidpunct}
{\mcitedefaultendpunct}{\mcitedefaultseppunct}\relax
\EndOfBibitem
\bibitem{Roe:2004na}
B.~P. Roe {\em et~al.}, \ifthenelse{\boolean{articletitles}}{\emph{{Boosted
  decision trees, an alternative to artificial neural networks}},
  }{}\href{https://doi.org/10.1016/j.nima.2004.12.018}{Nucl.\ Instrum.\ Meth.\
  A \textbf{543} (2005) 577}\relax
\mciteBstWouldAddEndPuncttrue
\mciteSetBstMidEndSepPunct{\mcitedefaultmidpunct}
{\mcitedefaultendpunct}{\mcitedefaultseppunct}\relax
\EndOfBibitem
\bibitem{LHCb-DP-2013-001}
F.~Archilli {\em et~al.},
  \ifthenelse{\boolean{articletitles}}{\emph{{Performance of the muon
  identification at LHCb}},
  }{}\href{https://doi.org/10.1088/1748-0221/8/10/P10020}{JINST \textbf{8}
  (2013) P10020}, \href{http://arxiv.org/abs/1306.0249}{{\normalfont\ttfamily
  arXiv:1306.0249}}\relax
\mciteBstWouldAddEndPuncttrue
\mciteSetBstMidEndSepPunct{\mcitedefaultmidpunct}
{\mcitedefaultendpunct}{\mcitedefaultseppunct}\relax
\EndOfBibitem
\bibitem{Dalitz:1953cp}
R.~H. Dalitz, \ifthenelse{\boolean{articletitles}}{\emph{{On the analysis of
  \Ptau-meson data and the nature of the \Ptau-meson}},
  }{}\href{https://doi.org/10.1080/14786441008520365}{Phil.\ Mag.\ Ser.\ 7
  \textbf{44} (1953) 1068}\relax
\mciteBstWouldAddEndPuncttrue
\mciteSetBstMidEndSepPunct{\mcitedefaultmidpunct}
{\mcitedefaultendpunct}{\mcitedefaultseppunct}\relax
\EndOfBibitem
\bibitem{Blatt:1952ije}
J.~M. Blatt and V.~F. Weisskopf, {\em {Theoretical nuclear physics}},
  \href{https://doi.org/10.1007/978-1-4612-9959-2}{ Springer, New York,
  1952}\relax
\mciteBstWouldAddEndPuncttrue
\mciteSetBstMidEndSepPunct{\mcitedefaultmidpunct}
{\mcitedefaultendpunct}{\mcitedefaultseppunct}\relax
\EndOfBibitem
\bibitem{VonHippel:1972fg}
F.~{von Hippel} and C.~Quigg,
  \ifthenelse{\boolean{articletitles}}{\emph{{Centrifugal-barrier effects in
  resonance partial decay widths, shapes, and production amplitudes}},
  }{}\href{https://doi.org/10.1103/PhysRevD.5.624}{Phys.\ Rev.\  \textbf{D5}
  (1972) 624}\relax
\mciteBstWouldAddEndPuncttrue
\mciteSetBstMidEndSepPunct{\mcitedefaultmidpunct}
{\mcitedefaultendpunct}{\mcitedefaultseppunct}\relax
\EndOfBibitem
\bibitem{LHCb-PAPER-2014-036}
LHCb collaboration, R.~Aaij {\em et~al.},
  \ifthenelse{\boolean{articletitles}}{\emph{{Dalitz plot analysis of
  \mbox{\decay{\Bs}{\Dzb\Km\pip}} decays}},
  }{}\href{https://doi.org/10.1103/PhysRevD.90.072003}{Phys.\ Rev.\
  \textbf{D90} (2014) 072003},
  \href{http://arxiv.org/abs/1407.7712}{{\normalfont\ttfamily
  arXiv:1407.7712}}\relax
\mciteBstWouldAddEndPuncttrue
\mciteSetBstMidEndSepPunct{\mcitedefaultmidpunct}
{\mcitedefaultendpunct}{\mcitedefaultseppunct}\relax
\EndOfBibitem
\bibitem{Aubert:2005ce}
BaBar collaboration, B.~Aubert {\em et~al.},
  \ifthenelse{\boolean{articletitles}}{\emph{{Dalitz-plot analysis of the
  decays $B^\pm \to K^\pm \pi^\mp \pi^\pm$}},
  }{}\href{https://doi.org/10.1103/PhysRevD.72.072003}{Phys.\ Rev.\
  \textbf{D72} (2005) 072003}, Erratum
  \href{https://doi.org/10.1103/PhysRevD.74.099903}{ibid.\   \textbf{74} (2006)
  099903}\relax
\mciteBstWouldAddEndPuncttrue
\mciteSetBstMidEndSepPunct{\mcitedefaultmidpunct}
{\mcitedefaultendpunct}{\mcitedefaultseppunct}\relax
\EndOfBibitem
\bibitem{Jackson:1964zd}
J.~D. Jackson, \ifthenelse{\boolean{articletitles}}{\emph{{Remarks on the
  phenomenological analysis of resonances}},
  }{}\href{https://doi.org/10.1007/BF02750563}{Nuovo Cim.\ \ \textbf{34} (1964)
  1644}\relax
\mciteBstWouldAddEndPuncttrue
\mciteSetBstMidEndSepPunct{\mcitedefaultmidpunct}
{\mcitedefaultendpunct}{\mcitedefaultseppunct}\relax
\EndOfBibitem
\bibitem{Back:2017zqt}
J.~Back {\em et~al.}, \ifthenelse{\boolean{articletitles}}{\emph{{{\sc
  Laura}$^{++}$: A Dalitz plot fitter}},
  }{}\href{https://doi.org/10.1016/j.cpc.2018.04.017}{Comput.\ Phys.\ Commun.\
  \textbf{231} (2018) 198},
  \href{http://arxiv.org/abs/1711.09854}{{\normalfont\ttfamily
  arXiv:1711.09854}}\relax
\mciteBstWouldAddEndPuncttrue
\mciteSetBstMidEndSepPunct{\mcitedefaultmidpunct}
{\mcitedefaultendpunct}{\mcitedefaultseppunct}\relax
\EndOfBibitem
\bibitem{Ben-Haim:2014afa}
E.~Ben-Haim, R.~Brun, B.~Echenard, and T.~E. Latham,
  \ifthenelse{\boolean{articletitles}}{\emph{{JFIT: a framework to obtain
  combined experimental results through joint fits}},
  }{}\href{http://arxiv.org/abs/1409.5080}{{\normalfont\ttfamily
  arXiv:1409.5080}}\relax
\mciteBstWouldAddEndPuncttrue
\mciteSetBstMidEndSepPunct{\mcitedefaultmidpunct}
{\mcitedefaultendpunct}{\mcitedefaultseppunct}\relax
\EndOfBibitem
\bibitem{Pelaez:2004vs}
J.~R. Pelaez and F.~J. Yndurain,
  \ifthenelse{\boolean{articletitles}}{\emph{{The pion-pion scattering
  amplitude}}, }{}\href{https://doi.org/10.1103/PhysRevD.71.074016}{Phys.\
  Rev.\  \textbf{D71} (2005) 074016}\relax
\mciteBstWouldAddEndPuncttrue
\mciteSetBstMidEndSepPunct{\mcitedefaultmidpunct}
{\mcitedefaultendpunct}{\mcitedefaultseppunct}\relax
\EndOfBibitem
\bibitem{Aaij:2019jaq}
LHCb collaboration, R.~Aaij {\em et~al.},
  \ifthenelse{\boolean{articletitles}}{\emph{{Amplitude analysis of the $B^+
  \rightarrow \pi^+\pi^+\pi^-$ decay}},
  }{}\href{http://arxiv.org/abs/1909.05212}{{\normalfont\ttfamily
  arXiv:1909.05212}}\relax
\mciteBstWouldAddEndPuncttrue
\mciteSetBstMidEndSepPunct{\mcitedefaultmidpunct}
{\mcitedefaultendpunct}{\mcitedefaultseppunct}\relax
\EndOfBibitem
\bibitem{Aaij:2019hzr}
LHCb, R.~Aaij {\em et~al.},
  \ifthenelse{\boolean{articletitles}}{\emph{{Observation of several sources of
  $CP$ violation in $B^+ \to \pi^+ \pi^+ \pi^-$ decays}},
  }{}\href{http://arxiv.org/abs/1909.05211}{{\normalfont\ttfamily
  arXiv:1909.05211}}\relax
\mciteBstWouldAddEndPuncttrue
\mciteSetBstMidEndSepPunct{\mcitedefaultmidpunct}
{\mcitedefaultendpunct}{\mcitedefaultseppunct}\relax
\EndOfBibitem
\bibitem{LHCb-PAPER-2014-013}
LHCb collaboration, R.~Aaij {\em et~al.},
  \ifthenelse{\boolean{articletitles}}{\emph{{Measurement of \CP asymmetry in
  \mbox{\decay{\Dz}{\Km\Kp}} and \mbox{\decay{\Dz}{\pim\pip}} decays}},
  }{}\href{https://doi.org/10.1007/JHEP07(2014)041}{JHEP \textbf{07} (2014)
  041}, \href{http://arxiv.org/abs/1405.2797}{{\normalfont\ttfamily
  arXiv:1405.2797}}\relax
\mciteBstWouldAddEndPuncttrue
\mciteSetBstMidEndSepPunct{\mcitedefaultmidpunct}
{\mcitedefaultendpunct}{\mcitedefaultseppunct}\relax
\EndOfBibitem
\end{mcitethebibliography}
